\documentclass[pdflatex,sn-nature]{sn-jnl}
\usepackage{graphicx}%
\usepackage{multirow}%
\usepackage{amsmath,amssymb,amsfonts}%
\usepackage{amsthm}%
\usepackage{mathrsfs}%
\usepackage[title]{appendix}%
\usepackage{xcolor}%
\usepackage{textcomp}%
\usepackage{manyfoot}%
\usepackage{booktabs}%

\usepackage{enumitem}
\usepackage{changepage}
\usepackage{pifont}

\begin{document}
\title{Emergent hidden order in ice: frustration and glassiness from slow hydrogen dynamics}

\author*[1]{\fnm{Tianran} \sur{Chen}}\email{tchen34@utk.edu}

\author[2]{\fnm{D. Jonathan P.} \sur{Morris}}\email{morrisd3@xavier.edu}

\author[1]{\fnm{Isaac C.} \sur{Ownby}}\email{iownby@vols.utk.edu}

\author[3]{\fnm{Anjana} \sur{Samarakoon}}\email{anjana8722@gmail.com}

\author[3,7]{\fnm{Arnab} \sur{Banerjee}}\email{arnabb@purdue.edu}

\author[3]{\fnm{Feng} \sur{Ye}}\email{yef1@ornl.gov}

\author[3]{\fnm{Douglas L.} \sur{Abernathy}}\email{abernathydl@ornl.gov}

\author[3]{\fnm{Zachary J.} \sur{Morgan}}\email{morganzj@ornl.gov}

\author[2,6]{\fnm{Joseph} \sur{Lanier}}\email{lanier.86@osu.edu}

\author[4]{\fnm{Konrad} \sur{Siemensmeyer}}\email{siemensmeyer@helmholtz-berlin.de}

\author[4]{\fnm{Bastian} \sur{Klemke}}\email{bastian.klemke@helmholtz-berlin.de}

\author*[1,5]{\fnm{D. Alan} \sur{Tennant}}\email{dtennant@utk.edu}

\affil*[1]{\orgdiv{Department of Physics and Astronomy}, \orgname{University of Tennessee}, \orgaddress{\street{1408 Circle Dr.}, \city{Knoxville}, \postcode{37996}, \state{TN}, \country{USA}}}

\affil[2]{\orgdiv{Department of Physics and Engineering}, \orgname{Xavier University}, \orgaddress{\street{3800 Victory Pkw.}, \city{Cincinnati}, \postcode{45207}, \state{OH}, \country{USA}}}

\affil[3]{\orgdiv{Oak Ridge National Laboratory}, \orgaddress{\street{1 Bethel Valley Rd.}, \city{Oak Ridge}, \postcode{37831}, \state{TN}, \country{USA}}}

\affil[4]{\orgdiv{Helmholtz-Zentrum Berlin für Materialien und Energie}, \orgaddress{\street{Hahn-Meitner-Platz 1}, \city{Berlin}, \postcode{14109}, \country{Germany}}}

\affil[5]{\orgdiv{Department of Materials Science and Engineering}, \orgname{University of Tennessee}, \orgaddress{\street{1001-1099 Estabrook Rd.}, \city{Knoxville}, \postcode{37996}, \state{TN}, \country{USA}}}

\affil[6]{\orgname{Department of Physics, The Ohio State University}, \orgaddress{\street{191 West Woodruff Ave.}, \city{Columbus}, \postcode{43210}, \state{OH}, \country{USA}}}

\affil[7]{\orgname{Department of Physics and Astronomy, Purdue University}, \orgaddress{\street{525 Northwestern Ave.}, \city{West Lafayette}, \postcode{47906}, \state{IN}, \country{USA}}}

\abstract{Frustrated systems can host hidden order, in which weak interactions select correlated structure from a highly degenerate manifold. Water ice Ih is the canonical example of such a manifold, yet whether its hydrogen disorder conceals local structure beyond the Bernal-Fowler ice rules has remained controversial. Here, using high-resolution inelastic neutron scattering on single-crystal heavy ice, we identify strongly anisotropic librational phonons dispersing uniaxially along the crystallographic \textit{c} axis — a spectroscopic signature inaccessible to bulk-averaging probes. A physics-guided analysis reveals that these excitations encode a hidden partial order: correlated polar armchair chains driven by a shallow stereochemical bias that creates an imperfectly flat energy landscape. This bias promotes nanoscale polar domains, yet frustrated topology prevents their straightforward coarsening into the ice XI ground state, trapping the system in a rugged configurational landscape that, within the experimentally constrained model, retains finite residual entropy even in the limit of infinitely slow cooling. These findings show that ice Ih is not a simple disordered solid, but a frustrated, partially ordered hydrogen network with glass-like arrest on a crystalline lattice, providing a microscopic framework that reconciles thermodynamic theory with spectroscopic observations. 
}


\maketitle

\section*{Main}\label{sec1} 

Hidden order, in which weak interactions select correlated structure from a highly degenerate manifold, is a recurring motif in frustrated matter, from spin ices to relaxor ferroelectrics \cite{samarakoon2022structural,samarakoon2022anomalous,bramwell2001spin,isakov2005spin,cross1987relaxor,kim2022coupled}. Directly identifying such hidden order in a bulk material, however, remains challenging. Doing so requires linking three ingredients within a single experimental framework: the microscopic interaction that biases the manifold, the correlated motif that emerges, and the mechanism that prevents the system from reaching its fully ordered ground state.

Water ice Ih provides a uniquely tractable arena for this problem. Its oxygen atoms form an ordered hexagonal lattice, while the hydrogen atoms remain disordered under the Bernal--Fowler ice rules, which require each oxygen atom to participate in two short covalent O--H bonds and two longer hydrogen bonds \cite{Bernal1933}. These local constraints generate a macroscopically degenerate manifold of allowed hydrogen configurations, giving rise to the residual configurational entropy predicted by Pauling \cite{Pauling1935} and confirmed experimentally by Giauque and Stout \cite{Giauque_Stout_1936}. At sufficiently low temperatures, a fully hydrogen-ordered ground state, ice XI, is known to exist \cite{Tajima1982,Fukazawa2006}. Yet hydrogen ordering has never been observed in pure ice Ih under laboratory conditions. This long-standing discrepancy raises a fundamental question: does the apparent disorder of ice Ih simply reflect the absence of long-range symmetry breaking, or does the hydrogen network harbor correlated organization hidden within the ice-rule manifold?

Previous studies have shown that hydrogen-disordered ice exhibits substantial correlations. In particular, diffuse scattering measurements have revealed pinch-point features characteristic of constrained disorder and long-range ice-rule correlations \cite{Li1994,Isakov2015}. However, these observations do not by themselves determine whether weaker stereochemical interactions further bias the manifold toward hidden local order. More generally, conventional crystallographic and thermodynamic probes average over the enormous configurational degeneracy of the hydrogen network, making it difficult to distinguish genuinely correlated motifs from statistical consequences of the ice rules alone. As a result, whether ice Ih hosts partial order beyond the Bernal--Fowler constraints has remained unresolved.

Here we address this question using high-resolution single-crystal inelastic neutron scattering (INS) on heavy ice (\mbox{$^2$H$_2$O}) \cite{Morris2019}. 
Because librational vibrations are exceptionally sensitive to hydrogen-bond geometry, they provide a spectroscopic probe of otherwise hidden orientational correlations that are difficult to access by bulk-averaging techniques. Mapping lattice dynamics across the full four-dimensional $(\mathbf{Q},E)$ space, we uncover strongly anisotropic librational phonons, including a higher-energy branch that disperses exclusively along the crystallographic $c$ axis. We show that this unusual uniaxial dispersion cannot be explained by the ice rules alone. A physics-guided inverse-modeling analysis links these excitations to hidden partial order in the form of correlated polar armchair chains selectively enhanced along the $c$ axis. This ordering is driven by a shallow stereochemical bias that lifts the ground-state degeneracy only partially, creating an imperfectly flat energy landscape where frustrated topology inhibits straightforward coarsening into the ice XI state. These findings provide a microscopic mechanism by which residual entropy can persist despite energetic bias toward hydrogen ordering, reconciling thermodynamic theory with spectroscopic observations and establishing ice as a tractable model system for glass-like arrest on a crystalline lattice.

\subsection*{Spectroscopic signature of hidden orientational order}\label{sec2}

\begin{figure}[t]
    \centering
    \includegraphics[page=1,width=\linewidth]{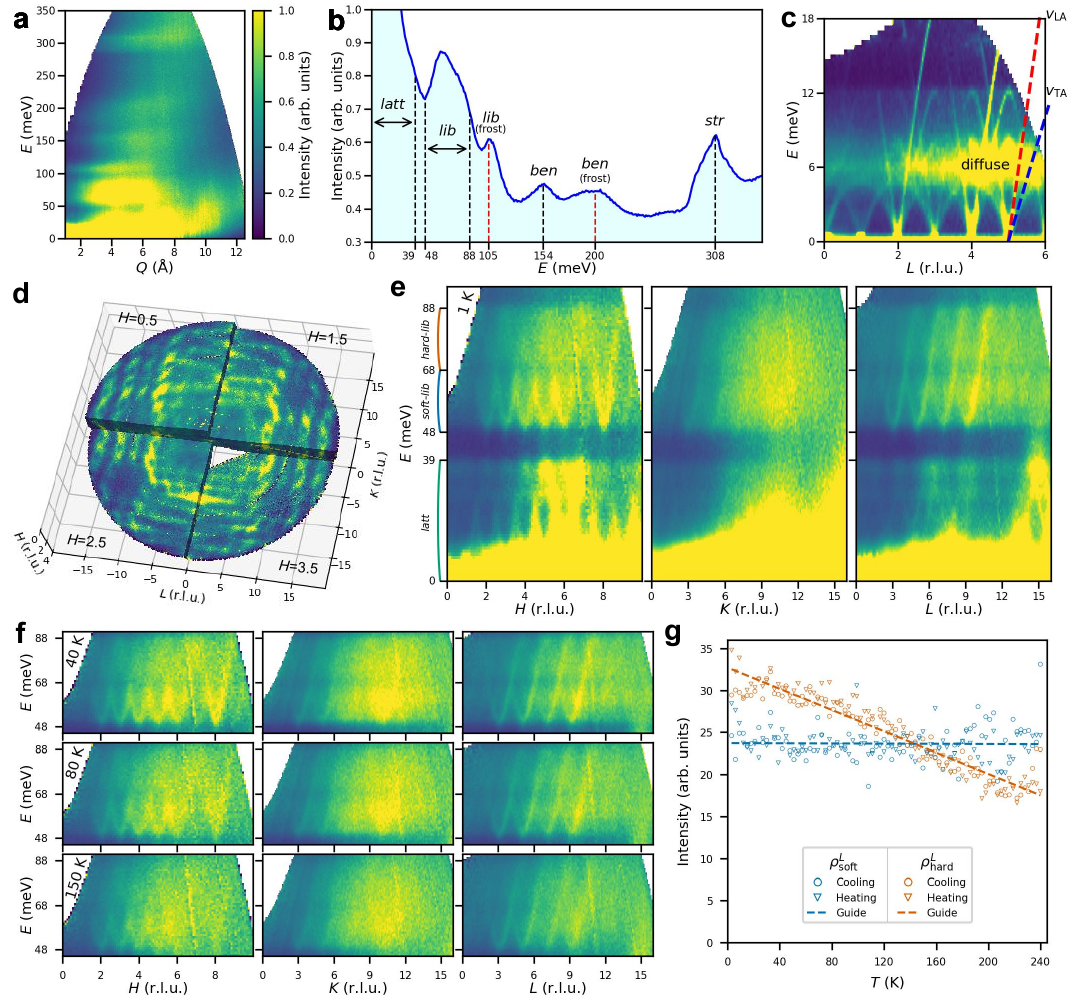}

\caption{\textbf{Anisotropic librational phonons in ice Ih.}
\textbf{a},~Inelastic neutron scattering (INS) intensity $I(\mathbf{Q},E)$ at 1~K. The colour scale applies to all neutron scattering data and simulations throughout.
\textbf{b},~Integrated spectrum $I(E)$ identifying four excitation groups: lattice (\textit{latt}, 0--39~meV), librational (\textit{lib}, 48--88~meV), bending (\textit{ben}, 154~meV), and stretching (\textit{str}, 308~meV). Minor peaks at 105 and 200~meV arise from surface $^1$H$_2$O frost (Supplementary Information).
\textbf{c},~Low-energy dispersion along $L$ resolving longitudinal (LA) and transverse (TA) acoustic phonons; slopes give group velocities $v_\text{LA}=3.8$~km~s$^{-1}$ and $v_\text{TA}=1.8$~km~s$^{-1}$. The broad continuum near 6~meV is thermal diffuse scattering.
\textbf{d},~$\mathbf{Q}$-dependent thermal diffuse scattering integrated over 4--12~meV at $H = [0.5, 1.5, 2.5, 3.5]$~r.l.u., selected to minimize acoustic phonon contributions.
\textbf{e},~Anisotropic librational dispersion at 1~K. 
\textit{Soft-lib} modes (48--68~meV) disperse along $H$ and $L$ but appear as broad continua along $K$; \textit{hard-lib} modes (68--88~meV) disperse exclusively along $L$.
\textbf{f},~Evolution of librational spectra at representative temperatures (40, 80, and 150~K).
\textbf{g},~Temperature-dependent dispersive spectral weights of \textit{soft-lib} (filled blue circles, heating; open blue circles, cooling) and \textit{hard-lib} (filled red triangles, heating; open red triangles, cooling) modes along $L$. Dashed lines are guides to the eye. Intensities are corrected for the Debye--Waller factor.}
\label{fig:1}
\end{figure}

The INS intensity $I(\mathbf{Q},E)$ at 1~K (Fig.~\ref{fig:1}a) and integrated spectrum $I(E)$ (Fig.~\ref{fig:1}b) identify four excitation groups: lattice (0--39~meV), librational (48--88~meV), bending (154~meV), and stretching (308~meV) modes. Low-energy spectra (Fig.~\ref{fig:1}c) resolve longitudinal and transverse acoustic phonons with group velocities $v_\text{LA}=3.8$~km s$^{-1}$ and $v_\text{TA}=1.8$~km s$^{-1}$, consistent with polycrystalline data \cite{Vogt2008}. A broad continuum near 6 meV appears as $\mathbf{Q}$-dependent thermal diffuse scattering (Fig.~\ref{fig:1}d), a signature of short-range hydrogen correlations.

The librational phonons (Fig.~\ref{fig:1}e) separate into a lower-energy branch (48--68 meV; hereafter \textit{soft-lib}) and a higher-energy branch (68--88 meV; hereafter \textit{hard-lib}). These modes exhibit pronounced anisotropy: \textit{soft-lib} modes disperse along $H$ and $L$ but appear as broad continua along $K$, while \textit{hard-lib} modes disperse only along $L$. This uniaxial dispersion of the \textit{hard-lib} branch is anomalous: as we will show later, it cannot be reproduced by any hydrogen configuration consistent with the ice rules alone, and constitutes the primary spectroscopic signature of hidden partial order. The coexistence of dispersive and non-dispersive components within each librational branch is consistent with partial ordering, with the dispersive spectral weight tracking the correlated fraction of the hydrogen network and the continuum reflecting the residual disordered fraction.

We tracked the evolution of these modes over a full thermal cycle (1~K $\leftrightarrow$ 240~K). Fig.~\ref{fig:1}f shows representative spectra. To quantify the temperature dependence, we define the dispersive spectral weight $\rho(T)$ as the momentum-dependent coherent intensity extracted after subtracting the smooth non-dispersive background (see Methods); the non-dispersive continuum, which reflects uncorrelated hydrogen disorder, is treated separately. The resulting $\rho(T)$ values (Fig.~\ref{fig:1}g) directly track the correlated fraction of the hydrogen network as a function of temperature.

Negligible hysteresis indicates that the observed correlations evolve reversibly over laboratory timescales, rather than reflecting purely frozen disorder. Of the components quantified in Fig.~\ref{fig:1}g, only $\rho^L_\text{hard}(T)$ shows strong temperature dependence, decreasing by $\sim$50\% upon heating, whereas $\rho^L_\text{soft}(T)$ remains nearly constant. The corresponding \textit{hard-lib} signal along $H$ is negligible in the spectra throughout the measured temperature range. This selectivity identifies the uniaxial \textit{hard-lib} dispersion as the unique spectroscopic channel sensitive to the hidden partial order.

The temperature-dependent spectral weight $\rho^L_\text{hard}(T)$ thus serves as an experimental proxy for the hidden partial order — a direct handle on correlations that bulk-averaging probes cannot access.

\subsection*{Physics-constrained inverse modeling of the energy landscape}\label{sec3}

\begin{figure}
    \centering
    \includegraphics[page=2,width=1\linewidth]{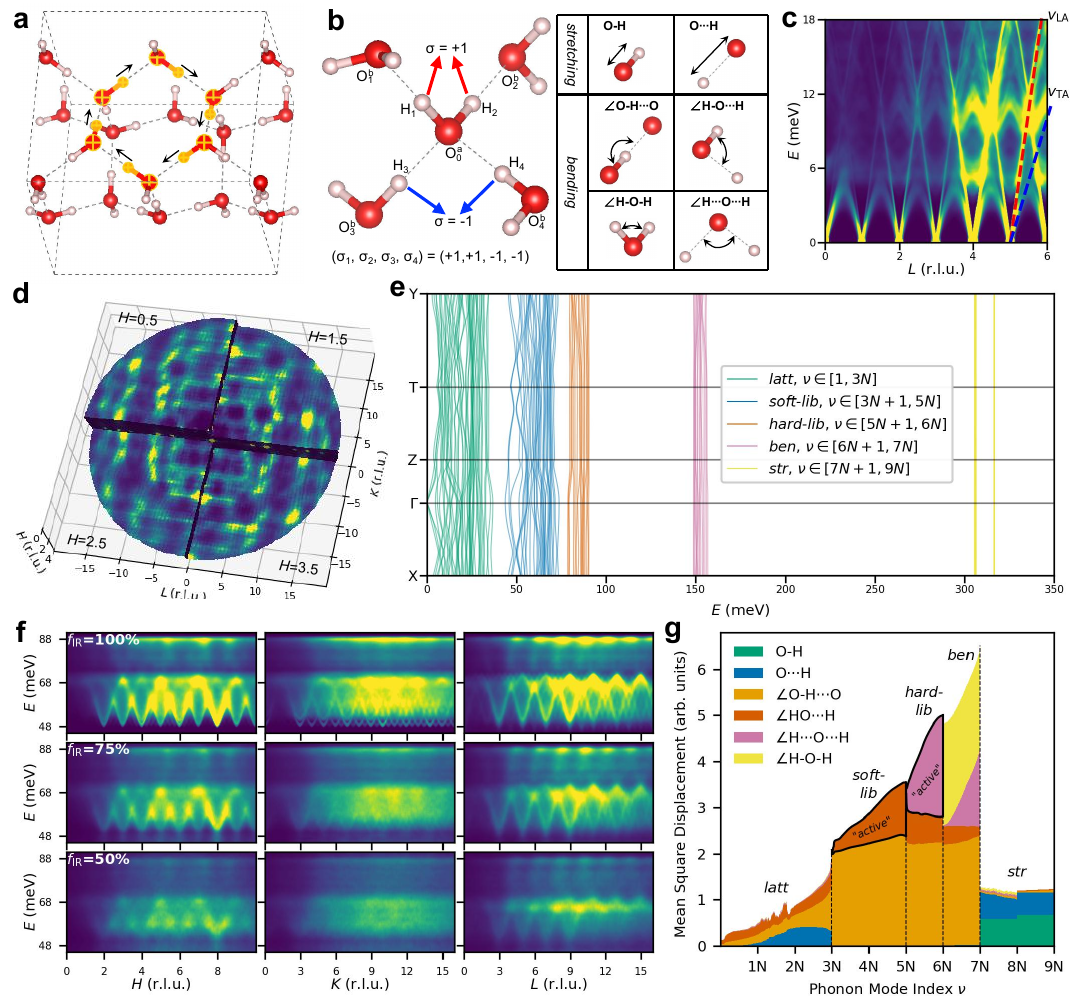}
\caption{\textbf{Physics-constrained inverse modeling of ice Ih lattice dynamics.}
\textbf{a},~Orthogonal conventional unit cell showing a closed ``O--H$\cdots$O--H$\cdots$'' loop; arrows indicate a collective loop-flip that preserves the ice rules.
\textbf{b},~Local hydrogen geometry for Ising configuration $(\sigma_1,\sigma_2,\sigma_3,\sigma_4)=(+1,+1,-1,-1)$, with two inequivalent oxygen sites O$^\text{a}$ (filled circles) and O$^\text{b}$ (open circles). The six harmonic force constants are indicated: bond-stretching ($k_\text{O-H}$, $k_{\text{O}\cdots \text{H}}$) and angle-bending ($k_{\angle\text{O-H}\cdots \text{O}}$, $k_{\angle\text{H-O}\cdots \text{H}}$, $k_{\angle\text{H-O-H}}$, $k_{\angle\text{H}\cdots \text{O}\cdots \text{H}}$).
\textbf{c},~Simulated one-phonon dynamic structure factor along $L$, reproducing the acoustic phonon dispersion of Fig.~1c.
\textbf{d},~Simulated multi-phonon thermal diffuse scattering, reproducing the $\mathbf{Q}$-dependent pattern of Fig.~1d.
\textbf{e},~Calculated phonon band structure along high-symmetry directions. Colours distinguish lattice (teal), \textit{soft-lib} (blue), \textit{hard-lib} (orange), bending (pink), and stretching (yellow) modes.
\textbf{f},~Simulated INS intensity along $H$, $K$, and $L$ for ice-rule compliance fractions $f_\text{IR} = 100\%$, $75\%$, and $50\%$ (top to bottom). Dispersive \textit{soft-lib} intensity scales with $f_\text{IR}$, while dispersive \textit{hard-lib} modes along $L$ are absent at all values, demonstrating their origin beyond the ice rules.
\textbf{g},~Mode-resolved mean-square displacements (MSDs) for all six force-constant coordinates as a function of phonon mode index $\nu$. Enhanced MSDs of $\angle$H-O$\cdots$H (dark orange) govern \textit{soft-lib} bandwidths; enhanced MSDs of $\angle$H$\cdots$O$\cdots$H (pink) govern \textit{hard-lib} bandwidths. Dashed vertical lines delimit the five mode families.}
\label{fig:2}
\end{figure}

Rather than fitting to a presumed microscopic theory, we employ a physics-constrained inverse modeling approach in which the experimental 4D phonon landscape directly constrains a minimal parameter set, allowing the energy landscape topology to be inferred from data. To implement this, we model the ice Ih lattice using an orthogonal conventional unit cell (Fig.~\ref{fig:2}a) containing two inequivalent oxygen sites, O$^{\mathrm{a}}$ and O$^{\mathrm{b}}$ (Fig.~\ref{fig:2}b), bridged by hydrogens displaced 0.35~\AA{} from the bond centers, forming covalent (O-H) and hydrogen (O$\cdots$H) bonds. We represent the hydrogen disorder by assigning an Ising pseudospin $\sigma_i=\pm 1$ to each hydrogen according to proximity to O$^{\mathrm{a}}$ or O$^{\mathrm{b}}$; the ice rules map to the local constraint $\sum_i \sigma_i = 0$ at each oxygen site (Fig.~\ref{fig:2}b).

Hydrogen configurations were generated using a loop-flip algorithm that performs collective hydrogen hopping along closed “O-H$\cdots$O-H$\cdots$” loops (Fig.~\ref{fig:2}a). This preserves the ice rules while enabling ergodic sampling of the allowed hydrogen-configuration manifold through cooperative loop updates \cite{presiado2011fast,Bove_PhysRevLett.103.165901,Nagle1966}.

Our minimal description is a six-parameter harmonic model containing two bond-stretching (O-H and O$\cdots$H) and four angle-bending ($\angle$O-H$\cdots$O, $\angle$H-O$\cdots$H, $\angle$H-O-H, and $\angle$H$\cdots$O$\cdots$H) force constants (Fig.~\ref{fig:2}b), fitted directly to the experimental phonon landscape. The potential is
\begin{equation}
V = \frac{1}{2}\sum_{\text{bonds}} k_b |\delta b|^2 + \frac{1}{2}\sum_{\text{angles}} k_\alpha |\delta \alpha|^2 ,
\end{equation}
where $\delta b$ and $\delta \alpha$ denote deviations in bond lengths and bond angles. 

This compact description reproduces the acoustic phonons, the $\mathbf{Q}$-dependent thermal diffuse scattering  (Figs.~\ref{fig:1}c-d and  \ref{fig:2}c-d), and the full phonon band structure (Figs.~\ref{fig:1}b and  \ref{fig:2}e), showing that the observed anisotropic dynamics can be captured by a compact set of local interactions.

The dispersive \textit{hard-lib} phonons along $L$ cannot be explained by the ice rules alone. Simulations enforcing only the ice rules reproduce the \textit{soft-lib} phonons but fail to generate the \textit{hard-lib} phonons (Fig.~\ref{fig:2}f). Systematically varying the fraction of ice-rule-satisfying sites,  $f_{\mathrm{IR}}$, confirms that \textit{soft-lib} intensity scales with $f_{\mathrm{IR}}$ while \textit{hard-lib} intensity remains negligible, demonstrating that the two mode families have distinct structural origins. The \textit{hard-lib} phonons therefore encode correlations beyond the ice rules: additional hydrogen ordering along $\hat{\mathbf{c}}$, selectively enhanced by an interaction not captured by the Bernal-Fowler constraints. The temperature-dependent suppression of $\rho^L_{\text{hard}}(T)$ (Fig.~\ref{fig:1}g) — with $\sim$50\% reduction upon heating and negligible hysteresis — confirms that these correlations are dynamic on laboratory timescales, accessible to slow cooling but not to nanosecond probes.

The unusually large librational bandwidth (48-88 meV) has been reported previously \cite{severson2003librational,wehinger2014diffuse,gupta2018phonons} but its microscopic origin has remained unclear. Its absence in \mbox{$^1$H$_2$O} frost (Fig.~\ref{fig:1}b) confirms it is a property of the crystalline hydrogen-bond network rather than local molecular environments. Mode-resolved mean-square displacement (MSD) analysis (Fig.~\ref{fig:2}g) reveals the origin: \textit{soft-lib} bandwidths arise from enhanced MSD of $\angle$H-O$\cdots$H, while \textit{hard-lib} bandwidths arise from $\angle$H$\cdots$O$\cdots$H — demonstrating that extended networks of these angular coordinates govern the librational breadth of ice Ih.

\subsection*{Stereochemical frustration selects correlated polar chains}\label{sec4}

\begin{figure}
    \centering
    \includegraphics[page=3,width=1\linewidth]{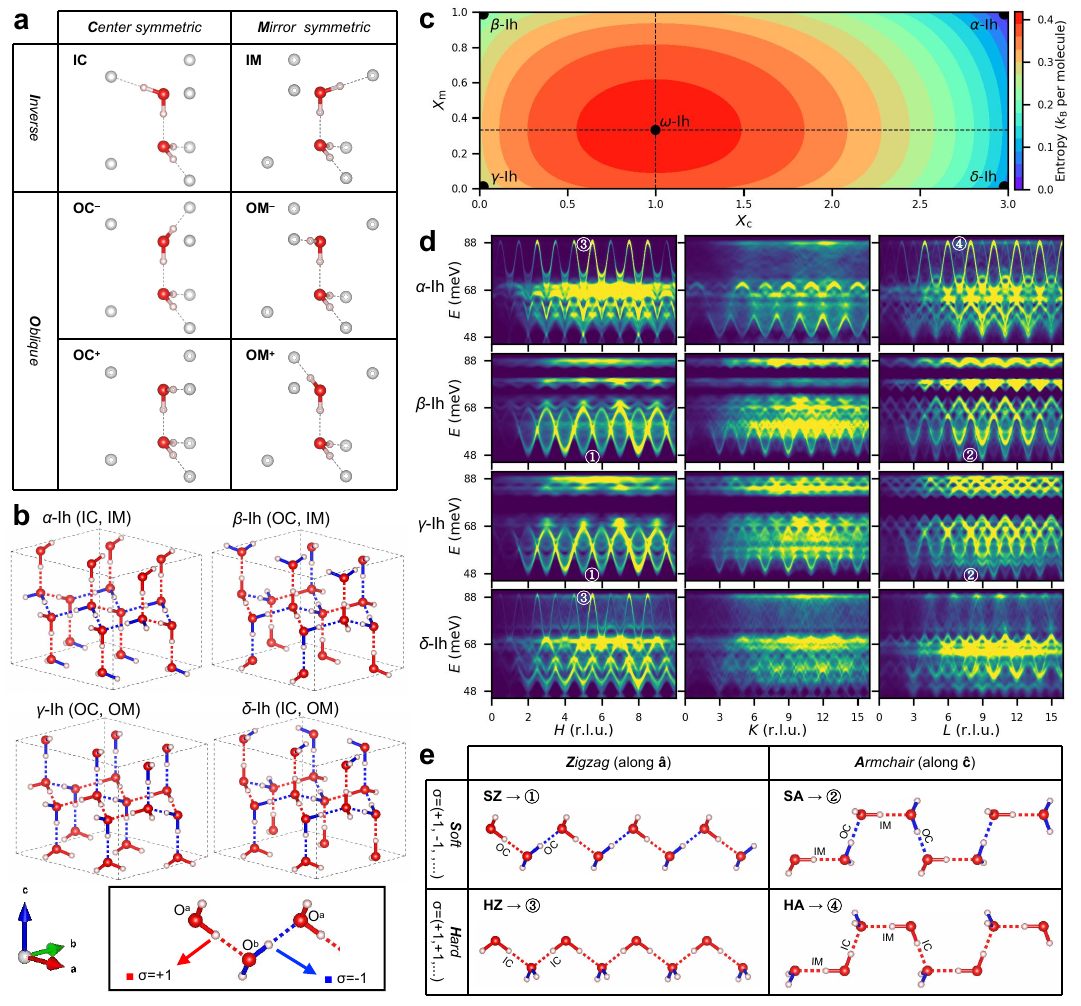}
\caption{\textbf{Stereochemical order parameter and hydrogen-bond chain classification.}
\textbf{a},~Six nearest-neighbor molecule-pair configurations in ice Ih, classified as center-symmetric (IC, OC$^\pm$) and mirror-symmetric (IM, OM$^\pm$).
\textbf{b},~Crystal structures of the four ordered phases ($\alpha$-, $\beta$-, $\gamma$-, and $\delta$-Ih), each defined by a specific pair-configuration combination. Hydrogen bonds with Ising variable $\sigma=+1$ are shown in red; those with $\sigma=-1$ are shown in blue. The inset identifies the two inequivalent oxygen sites O$^\text{a}$ (filled circles) and O$^\text{b}$ (open circles).
\textbf{c},~Conformational entropy $S(X_\text{c},X_\text{m})$ as a function of order parameter. The maximally disordered $\omega$-Ih state (filled circle) lies at $\mathbf{X}=(1,\,1/3)$ with $S=k_\text{B}\ln(3/2)$; the four ordered phases occupy the corners. Colour scale indicates entropy in units of $k_\text{B}$ per molecule, from zero (dark blue, fully ordered) to $k_\text{B}\ln(3/2)$ (dark red, maximally disordered).
\textbf{d},~Simulated librational phonon spectra of the four ordered phases along $H$, $K$, and $L$. Four distinct dispersive modes are identified (circled numerals 1--4): \textit{soft-lib} along $H$ (mode 1) and $L$ (mode 2), and \textit{hard-lib} along $H$ (mode 3) and $L$ (mode 4).
\textbf{e},~Four hydrogen-bond chain types — soft-zigzag (SZ), soft-armchair (SA), hard-zigzag (HZ), and hard-armchair (HA) — corresponding to dispersive modes 1--4 in \textbf{d}. Zigzag chains propagate along $\mathbf{\hat{a}}$; armchair chains propagate along $\mathbf{\hat{c}}$. The constituent pair symmetries (IC, OC, IM, OM) are indicated for each chain. HA-chains carry longitudinal polarization along $\mathbf{\hat{c}}$ (indicated by the uniform Ising sequence), making them the structural unit of nanoscale polar domains.}
\label{fig:3}
\end{figure}

The experimental observation of dispersive \textit{hard-lib} phonons along $L$ implies the presence of hydrogen correlations beyond the ice rules. The key point of the classification introduced below is that the experimentally observed hidden order corresponds specifically to enhanced hard-armchair (HA) chains along the crystallographic $c$ axis. To identify the stereochemical origin of this bias, we classify nearest-neighbor water molecule pairs by geometric symmetry (Fig.~\ref{fig:3}a): center-symmetric (inverse, IC; or oblique, OC$^\pm$) and mirror-symmetric (inverse, IM; or oblique, OM$^\pm$) \cite{bjerrum1952structure}. Within each symmetry class, configurations can interconvert via 3-fold rotations, but center- and mirror-symmetric pairs cannot. Each molecule participates in four such pairs: three center-symmetric and one mirror-symmetric. We define an order parameter $\mathbf{X}=(X_\text{c},X_\text{m})$ as the average number of IC and IM pairs per molecule ($X_\text{c}\in[0,3]$, $X_\text{m}\in[0,1]$), yielding four ordered states (Fig.~\ref{fig:3}b): $\alpha$-Ih ($\mathbf{X}=(3,1)$), $\beta$-Ih $(0,1)$, $\gamma$-Ih $(0,0)$, and $\delta$-Ih $(3,0)$.

The conformational entropy per molecule is estimated as a function of $\mathbf{X}$ (Fig.~\ref{fig:3}c; see Supplementary Information):
\begin{equation}
S(\mathbf{X}) \approx k_\text{B} \frac{\ln(3/2)}{4 \ln 3} \Biggl[
\phi(3)-\phi(X_\text{c})-\phi(X_\text{m})
- 2\phi\Bigl(\frac{3-X_\text{c}}{2}\Bigr)
-2 \phi\Bigl(\frac{1-X_\text{m}}{2}\Bigr)
\Biggr],
\end{equation}
where $k_\text{B}$ is Boltzmann's constant and $\phi(x) = x \ln x$. The $\omega$-Ih state at $\mathbf{X}=(1, 1/3)$ corresponds to the maximally disordered configuration, with $S_{\omega\text{-Ih}} = k_\text{B} \ln(3/2)$, consistent with Pauling's prediction \cite{Pauling1935}. The four ordered structures define distinct local minima in the conformational entropy landscape, separated by collective hydrogen reconfiguration barriers; $\alpha$-Ih is the ice XI ground state.

Phonon spectra of the four ordered structures reveal four distinct dispersive librational modes (Fig.~\ref{fig:3}d, circled numeral 1-4), each mapping directly to a hydrogen-bond chain type (Fig.~\ref{fig:3}e) classified by stiffness (soft or hard, S/H) and geometry (zigzag along $\mathbf{\hat{a}}$ or armchair along $\mathbf{\hat{c}}$, Z/A). Soft chains (SZ, SA) follow ``O-H$\cdots$O-H$\cdots$'' with alternating Ising sequence $\{\sigma\}_{\text{soft}}=(+1,-1,\dots)$ and weaker stiffness $k_{\angle\text{H-O$\cdots$H}}$. Hard chains (HZ, HA) follow ``H-O-H$\cdots$O$\cdots$'' with uniform Ising sequence $\{\sigma\}_{\text{hard}}=(+1,+1,\dots)$ and stronger stiffness  $k_{\angle\text{H$\cdots$O$\cdots$H}}$ (consistent with the MSD analysis, Fig.~\ref{fig:2}g). Critically, the uniform Ising sequence of hard chains imparts a local net polarization: HA-chains carry longitudinal polarization along $\mathbf{\hat{c}}$, while HZ-chains carry transverse polarization along $\mathbf{\hat{a}}$. 

The dispersive \textit{soft-lib} phonons along $H$ and $L$ originate from SZ- and SA-chains; \textit{hard-lib} phonons along $H$ and $L$ originate from HZ- and HA-chains. Zigzag and armchair chains are the only geometrically planar chains in ice Ih, facilitating coherent phonon propagation, whereas twisted chains hinder dispersion. Dispersive librational phonons along $K$ are largely suppressed because no chains directly align with $\mathbf{\hat{b}}$.

Crucially, the ice rules preferentially enhance \textit{soft-lib} phonons while suppressing \textit{hard-lib} phonons. Enforcing the ice rules increases the probability of opposite-sign Ising pairs ((+1,-1), from 1/2 to 2/3) while decreasing same-sign pairs ((+1,+1), from 1/4 to 1/6) — statistically favoring SZ/SA-chains and suppressing HZ/HA-chains. Consequently, ice-rule-constrained simulations (Fig.~\ref{fig:2}f) reproduce the \textit{soft-lib} phonons but fail to generate the \textit{hard-lib} phonons along $L$. The experimental observation of dispersive \textit{hard-lib} modes along $L$ is therefore evidence most consistent with an ordering mechanism beyond the ice rules: a stereochemical bias that selectively promotes HA-chains along $\mathbf{\hat{c}}$, associated with specific IC/OC and IM/OM pair configurations (Fig.~\ref{fig:3}e). We quantify this bias in the following section.

\subsection*{Intrinsic glassiness from an imperfectly flat energy landscape}\label{sec5}

Our analysis identifies HA-chains as the spectroscopic signature of hidden order, but leaves open the quantitative question: what energetic bias selects them, and can it be extracted from experiment? In ice Ih, direct quantification of stereochemical energies has proven elusive — the six molecule-pair configurations (Fig.~\ref{fig:3}a) are chemically indistinguishable and rapidly fluctuating, yielding only bulk-averaged observables. Here we invert this problem: rather than inferring these energies from theory, we extract an effective stereochemical bias by jointly fitting the experimental phonon and diffuse-scattering data within the minimal model. We model the internal energy per molecule as a function of the order parameter $\mathbf{X}$:
\begin{equation}
  U(\mathbf{X}) = -\frac{1}{2}(X_{\text{c}}\Delta_{\text{c}} + X_{\text{m}}\Delta_{\text{m}}),
\end{equation}
where $\Delta_{\text{c}} = E_{\text{OC}} - E_{\text{IC}}$ and $\Delta_{\text{m}} = E_{\text{OM}} - E_{\text{IM}}$. We express $\mathbf{\Delta}=(\Delta_\text{c},\Delta_\text{m})$ in polar coordinates $(\Delta,\theta)$ for compactness.

We performed Markov-chain Monte Carlo simulations using collective loop-flip dynamics (Fig.~\ref{fig:2}a) and the Metropolis criterion, cooling from the disordered $\omega$-Ih manifold ($T= 3\Delta$) down to $T=0$. The trajectory in the order-parameter plane $\mathbf{X}$ depends on the energetic anisotropy $\theta$ and the dimensionless sampling density $\Gamma$, defined as the number of attempted loop-flips per unit cell per temperature interval $\Delta$.

\begin{figure}
    \centering
    \includegraphics[page=4,width=\linewidth]{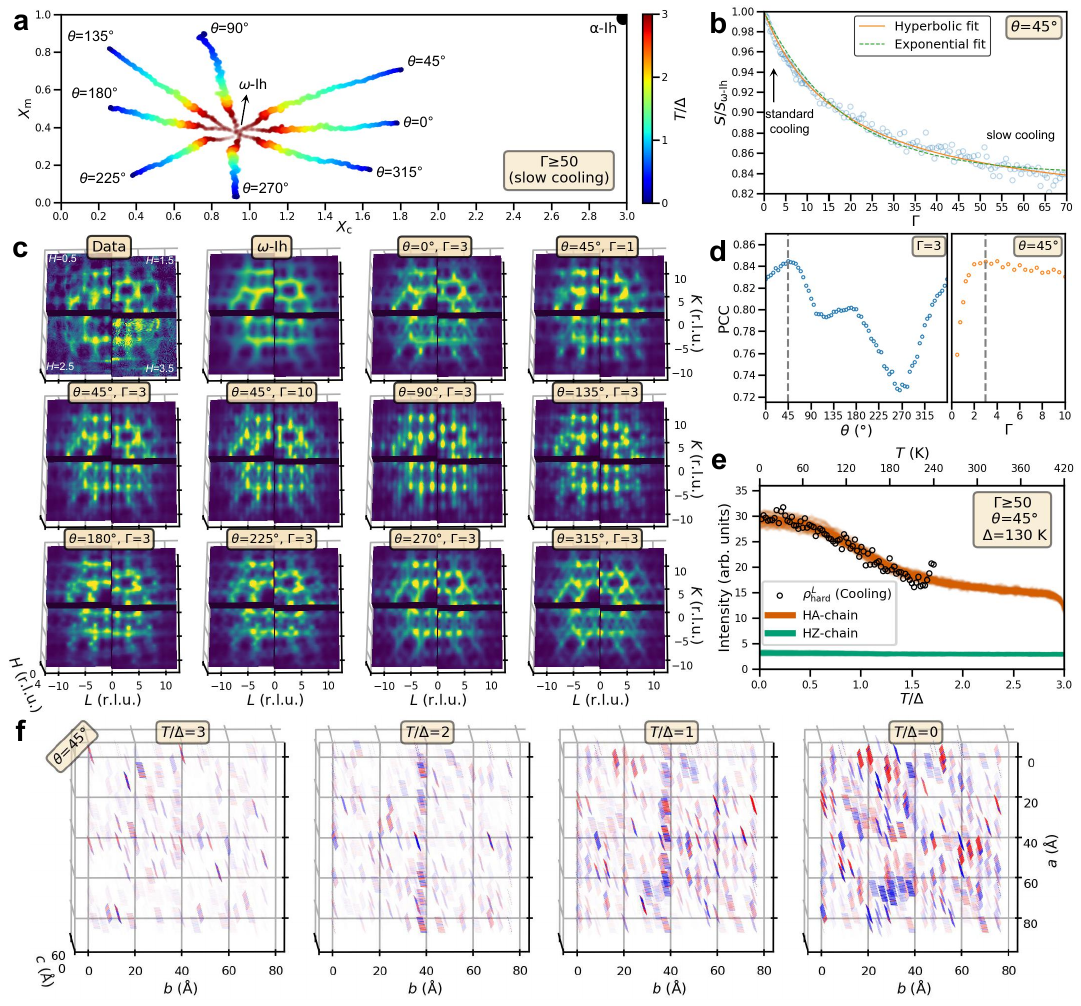}
\caption{\textbf{Energetic and kinetic constraints on stereochemical bias and domain formation.}
\textbf{a},~Monte Carlo cooling trajectories in the order-parameter plane $\mathbf{X}$ for $20\times10\times10$ supercell ($\Gamma\geq50$), coloured from red (high $T/\Delta$) to blue (low $T/\Delta$). Each trajectory corresponds to a different $\theta$, evolving from $\omega$-Ih toward a distinct partially ordered state, with $\alpha$-Ih selected near $\theta\approx45^\circ$.
\textbf{b},~Final-state entropy $S/S_{\omega\text{-Ih}}$ versus $\Gamma$ at $\theta=45^\circ$. Open circles are simulation data; solid orange and dashed green lines show the best hyperbolic and exponential fits, respectively. Vertical arrows mark the standard-cooling and slow-cooling regimes. Error bars indicate 
standard deviation.
\textbf{c},~Experimental elastic diffuse scattering at 1~K (top left) compared with simulated structure factors for the $\omega$-Ih state and representative $(\theta,\Gamma)$ combinations. The $\theta=45^\circ$, $\Gamma=3$ simulation best reproduces the experimental pattern.
\textbf{d},~Pearson correlation coefficient (PCC) between simulated and experimental structure factors versus $\theta$ (left, fixed $\Gamma=3$) and versus $\Gamma$ (right, fixed $\theta=45^\circ$). Dashed vertical lines mark optimal values.
\textbf{e},~Simulated HA-chain (orange) and HZ-chain (green) intensities versus $T/\Delta$ (bottom) and $T$ (top) for $\theta=45^\circ$, $\Gamma\geq50$, $\Delta=130$~K. Open circles show experimental $\rho^L_\text{hard}(T)$ (from Fig.~1g). HA-chain intensity reproduces the experimental profile; HZ-chain intensity is negligible throughout.
\textbf{f},~Snapshots of HA-chain configurations at $T/\Delta=3$, $2$, $1$, $0$ for $\theta=45^\circ$. Chains with $\sigma=+1$ are shown in solid red; $\sigma=-1$ in solid blue; mixed-sequence chains are transparent. Domains grow anisotropically along $\hat{\mathbf{c}}$, exceeding 80~\AA{} at $T/\Delta=0$.}
\label{fig:4}
\end{figure}

In the limit of rapid equilibration ($\Gamma\geq 50$; Fig.~\ref{fig:4}a), the trajectory is uniquely controlled by $\theta$, directing the system toward a partially ordered ground state. Finite kinetics intercept this cooling path: for a given $\theta$ (here exemplified as 45$^\circ$), the final configurational entropy follows an empirically identified hyperbolic decay (Fig.~\ref{fig:4}b):
\begin{equation}
S/S_{\omega\text{-Ih}}=0.80+\frac{0.20}{1+\Gamma/18.8}.
\end{equation}
This functional form — identified from simulation data rather than derived analytically — captures the cooperative nature of the ordering process more accurately than a simple exponential relaxation. 
Within the present model, extrapolating to $\Gamma \rightarrow \infty$ yields a residual entropy floor of $0.80 S_{\omega\text{-Ih}}$ — distinct from both the Pauling state and ice XI — an intermediate thermodynamic regime and the quantitative signature of intrinsic glassiness: disorder that persists not because cooling is too fast, but because the energy landscape itself is rugged.
 
Two independent experimental datasets jointly constrain the model parameters. The elastic diffuse scattering at 1~K (Fig.~\ref{fig:4}c), collected at a standard cooling rate  ($\sim 3$~K~min$^{-1}$), constrains $\Gamma$ and $\theta$: the best agreement occurs for $\theta = 45^\circ \pm 15^\circ$ and $\Gamma \gtrsim 3$ (Fig.~\ref{fig:4}d), with asymmetric sensitivity in $\Gamma$  that strictly bounds the lower limit. The slow-cooled ($ 0.15$~K~min$^{-1}$), $\rho^L_{\text{hard}}(T)$ dispersive spectral weight independently constrains $\Delta $: fitting the simulated HA-chain intensity ($\theta = 45^\circ$, $\Gamma\geq 50$) yields $\Delta = 130 \pm 15$~K (Fig.~\ref{fig:4}e). 

The kinetic mapping $A \approx \Gamma R_\text{cool}/\Delta$ translates the experimental cooling-rate constraint directly into an effective collective attempt frequency: $A \gtrsim 4$ loop-flips per unit cell per hour. This provides a rare experimental bridge between macroscopic thermodynamic observables and the effective timescale of collective hydrogen reconfiguration — without requiring molecular dynamics simulation or single-molecule spectroscopy.

The experimentally determined $\mathbf{\Delta}$ contrasts sharply with classical electrostatic predictions on two independent grounds. First, Bjerrum's point-charge model predicts $\Delta = 530\text{--}730$~K — four times larger than our extracted value — implying effectively frozen hydrogen configurations at experimentally accessible temperatures, in direct conflict with the observed temperature dependence of $\rho^L_{\text{hard}}(T)$. Second, the predicted orientation $\theta = 311\text{--}326^\circ$ favors $\delta$-Ih as the lowest-energy state, contradicting the experimentally established ice XI ground state. Our extracted parameters ($\theta \approx 45^\circ$, $\Delta \approx 130$~K) resolve both failures simultaneously: $\alpha$-Ih (ice XI) is correctly placed as the global minimum, and hydrogen reconfiguration remains dynamically accessible to cryogenic temperatures, demonstrating that short-range stereochemical interactions beyond classical electrostatics govern the true energy landscape of ice Ih.

Snapshots of the simulated cooling trajectory (Fig.~\ref{fig:4}f) reveal the kinetic topology of the resulting state. While HZ-chains form along six symmetry-equivalent directions in the ab-plane, polarized HA-chains emerge exclusively along $\pm\mathbf{\hat{c}}$. This selectivity enables efficient HA-chain merging along $\mathbf{\hat{c}}$, producing anisotropic polar domains ($> 80$~\AA{} along $\mathbf{\hat{c}}$ versus $10\text{--}20$~\AA{} in the $ab$ plane) that locally resemble ice XI. These nanoscale polar regions — with local polarization along $\pm\mathbf{\hat{c}}$ — constitute a direct, falsifiable prediction: they should be detectable by second-harmonic generation microscopy or by anomalies in the dielectric susceptibility at cryogenic temperatures \cite{Denev2011, Rick2003dielectric}. We emphasize that these domains arise from short-range stereochemistry rather than long-range electrostatics, distinguishing them from the KOH-doping-induced long-range ferroelectric order of ice XI \cite{Tajima1982}.

\section*{Discussion}\label{sec6}

The dispersive \textit{hard-lib} phonons along $L$ provide spectroscopic evidence for hidden partial order in ice Ih: correlated polar HA-chains along $\hat{\mathbf{c}}$, selected by a shallow stereochemical bias within the ice-rule-constrained manifold. This hidden order is not simply a consequence of finite cooling rate; rather, it reflects the topology of the underlying energy landscape. The experimentally constrained model predicts that, even in the limit of infinitely slow cooling, the system does not reach the ice XI ground state but instead converges to a finite residual entropy floor—a quantitative signature of intrinsic glassiness.

These findings reconcile thermodynamic theory with historical measurements of residual entropy. The classic confirmation of Pauling's entropy \cite{Giauque_Stout_1936} relied on standard cooling rates ($\sim 3$~K~min$^{-1}$), which our simulations confirm lead to kinetic arrest in a nearly disordered state ($S \approx 0.97\,S_{\omega\text{-Ih}}$). More fundamentally, the experimentally constrained model predicts that even at $\Gamma \to \infty$, the system converges to $\sim 0.80 S_{\omega\text{-Ih}}$ — distinct from both the Pauling state and ice XI. 

Mechanistically, this metastability reflects a competition between two opposing tendencies. The ice rules entropically suppress HA-chain formation, while the stereochemical bias $\mathbf{\Delta}$ energetically favors the IC/IM configurations that constitute them. As temperature decreases, the energetic term progressively wins: HA-chain density grows, nucleating nanoscale polar domains with opposing local polarizations ($\pm\hat{\mathbf{c}}$). But coarsening is arrested by a second frustration: as domains impinge, boundary frustration locks the system into a glassy mosaic — flipping an entire domain requires cooperative rearrangements prohibitively costly for short-range interactions alone. It is this interplay — entropic suppression at the chain level, geometric frustration at the domain level — that produces the finite residual entropy floor, directly analogous to spin-ice systems \cite{samarakoon2022structural,samarakoon2022anomalous} and entropy-stabilized phases in pyrochlore magnets \cite{Isakov2015}. Ice Ih thus exemplifies hidden order selected by energy landscape topology rather than symmetry breaking, a mechanism with clear analogues across spin, orbital, and structural degrees of freedom. More broadly, the work shows that anisotropic lattice dynamics can serve as a spectroscopic probe of hidden local correlations that are inaccessible to conventional bulk-averaging measurements.

Ice Ih offers a uniquely tractable model system for glass-like arrest: described by nine parameters — six harmonic force constants, two stereochemical energy parameters, and one kinetic attempt frequency — fitted directly to experiment rather than derived from theory. This physics-constrained inverse modeling approach, analogous in spirit to physics-informed machine learning but grounded in interpretable parameters, reproduces the full complexity of the 4D phonon landscape and directly connects nanosecond-scale molecular dynamics to laboratory cooling rates. The parsimony is itself a physical result: a rare case where atomic-scale interactions, mesoscale domain structure, and macroscopic thermodynamic arrest are connected quantitatively within a single experimentally constrained framework.

The domain-coarsening mechanism offers a natural explanation for the geological timescale of ice XI formation \cite{Fukazawa2006}: escaping the stereochemical trap requires overcoming boundary frustration through weaker, long-range electrostatic fields — consistent with the known requirement for KOH doping to achieve ice XI under laboratory conditions \cite{Tajima1982}. Such fields could bias domain coarsening over geological timescales, explaining why ice Ih persists as a distinct phase. This has direct implications for icy moon interiors such as Europa and Enceladus, where planetary-timescale cooling may produce a partially ordered hydrogen network detectable through anomalies in dielectric response or seismic velocity.

\backmatter

\clearpage

\clearpage
\clearpage

\clearpage
\clearpage

\section*{Methods}\label{sec:methods}

\subsection*{Sample preparation}

Heavy water (\mbox{$^2$H$_2$O}) was used throughout to minimize incoherent neutron scattering background \cite{Sears1992}. Single-crystal samples (12~cm length, 1~cm diameter) were grown at the Helmholtz-Zentrum Berlin using a modified Bridgman technique \cite{Ohtomo1987, Morris2019}; full growth protocol details are provided in the Supplementary Information. Single-crystallinity was verified by optical birefringence prior to neutron scattering measurements.

\subsection*{Sample characterization}

Minor peaks at 105 and 200~meV in the integrated INS spectrum arise from surface $^1$H$_2$O frost contamination, estimated at 2--3\% by comparing the incoherent scattering intensities of the $^2$H$_2$O bending mode (154~meV) and the $^1$H$_2$O bending mode (200~meV) using tabulated cross-sections ($\sigma_\text{inc}$: $^2$H = 2.05~barn, $^1$H = 80.27~barn). The frost is identified as surface $^1$H$_2$O rather than HDO by the absence of a peak at 185~meV. This contamination does not affect the dispersive librational features, because $^1$H contributes only weakly to the coherent scattering signal.

\subsection*{Neutron scattering experiments}

Elastic and inelastic neutron scattering measurements were performed at the Spallation Neutron Source (SNS), Oak Ridge National Laboratory. The sample was mounted under a 300~mbar helium atmosphere to suppress sublimation; the cryostat was purged with helium gas at 190~K prior to sample loading to prevent condensation.

\textbf{Elastic diffuse scattering.} Elastic scattering was measured on the CORELLI spectrometer \cite{CORELLI}, using a cross-correlation chopper to discriminate elastic from inelastic scattering. The sample was rotated in 3$^\circ$ steps from 180$^\circ$ to 360$^\circ$, with 4-minute dwell times per angle. The 1~K dataset was collected following a standard instrument cooling protocol ($\sim 3$~K~min$^{-1}$).

\textbf{Inelastic neutron scattering.} Following CORELLI measurements, the cryostat was transferred to the Wide Angular-Range Chopper Spectrometer (ARCS) \cite{ARCS}. Phonon dispersions were measured using the high-resolution Fermi chopper ``ARCS-100-1.5'' spinning at 420~Hz (native resolution $\approx 4\%$ of $E_i$). Incident energies were set to $E_i = 20$~meV (lattice modes), 150~meV (librational and bending modes), and 400~meV (stretching modes). The sample was rotated by 2$^\circ$ every 4~minutes. Temperature-dependent measurements used a slow-cooling protocol ($\sim 0.15$~K~min$^{-1}$) from 240~K to 1~K, with 20-minute dwell times at each 3~K step.

\subsection*{Analysis of dispersive librational phonons}\label{sec:data_analysis}

We quantified dispersive librational phonon intensities by integrating the four-dimensional scattering intensity \(I(H,K,L,E)\). The integrated intensity as a function of \(L\) is defined as:
\[
I(L) = \int_{-2}^{2} dH \int_{-2}^{2} dK \int_{E^-}^{E^+} dE \; I(H, K, L, E + \mu L),
\]
where \((E^-, E^+)\) are (48, 68)~meV for \textit{soft-lib} modes and (68, 88)~meV for \textit{hard-lib} modes, and \(\mu = 20~\text{meV/r.l.u.}\) accounts for the branch slope (tracking the dispersive ridge). To remove the smooth background, we performed an analogous integration along \(K\) (where no dispersion is observed) to estimate the smooth background, $I_\text{bg}(L)$, and subtracted it: \(I_{\text{peak}}(L) = I(L) - I_\text{bg}(L)\). The resulting peaks were integrated to obtain intensities \(\rho^L(i,T)\) for the $i$-th peak.

Thermal attenuation was corrected using the Debye-Waller factor $\exp(-Q^2\langle u^2\rangle/2)$, with mean-square displacement \(\langle u^2 \rangle = 5.5 \times 10^{-5}~\text{\AA}^2\text{K}^{-1} \cdot T\) determined by fitting. The corrected intensities (denoted $\rho$ throughout the main text) collapsed onto a single temperature-independent curve at low temperature, confirming the robustness of the correction.

Quantitative temperature-dependent spectral weights were extracted only along $L$; the absence of a corresponding dispersive \textit{hard-lib} signal along $H$ was assessed qualitatively from the temperature-dependent spectra.

\subsection*{Phonon structure factor calculations}\label{sec:DSF_simulation}

The one-phonon dynamic structure factor was computed as \cite{SquiresBook}:
\[
S(\mathbf{Q},\nu,\omega)=
\frac{k'}{k} N \,
\mathrm{e}^{-\langle|\mathbf{Q}\cdot\mathbf{u}|^2\rangle}
\sum_{\mathbf{q}=\mathbf{Q}}
\left|
\sum_{i}
\frac{\bar{b}_{i}}{\sqrt{2 m_i \omega_{\mathbf{q}\nu}}}
\left( \mathbf{Q}\cdot\mathbf{e}^{\,i}_{\mathbf{q}\nu} \right)
\mathrm{e}^{\mathrm{i} \mathbf{Q}\cdot\mathbf{r}_{i}}
\right|^{2}
\bigl(n_{\mathbf{q}\nu}+1 \bigr)
\delta(\omega-\omega_{\mathbf{q}\nu}),
\]
where $\mathbf{k}$ and $\mathbf{k'}$ are the initial and final neutron wavevectors, $\mathbf{Q}=\mathbf{k'}-\mathbf{k}$ is the momentum transfer, $N$ is the number of atoms, $\bar{b}_i$ and $m_i$ are the coherent scattering length and mass of atom $i$, $\mathbf{r}_i$ is the equilibrium position, and $\mathbf{e}^i_{\mathbf{q}\nu}$ is the phonon polarization vector for mode $(\mathbf{q},\nu)$. $e^{-\langle |\mathbf{Q}\cdot\mathbf{u}|^2\rangle}$ is the Debye-Waller factor, and $n_{\mathbf{q}\nu}$ is the phonon number given by the Bose-Einstein distribution. 

For coherent multi-phonon scattering, we used the corresponding expression \cite{SquiresBook}:
\[
S(\mathbf{Q},\omega)=
\frac{k'}{k} N \,
\mathrm{e}^{-\langle|\mathbf{Q}\cdot\mathbf{u}|^2\rangle}
\sum_{\mathbf{q}}
\sum_{\omega_{\mathbf{q}\nu}\le\omega}
\int \! dt \,
\left|
\sum_{i}
\frac{\bar{b}_i}{\sqrt{2 m_i \omega_{\mathbf{q}\nu}}}
\mathrm{e}^{\mathrm{i}\mathbf{Q}\cdot\bigl(\mathbf{r}_{i} + \mathbf{u}^i_{\mathbf{q}\nu}(t)\bigr)}
\right|^{2}
\bigl(n_{\mathbf{q}\nu}+1 \bigr),
\]
where $\mathbf{u}^i_{\mathbf{q}\nu}(t)$ is the instantaneous displacement associated with mode $(\mathbf{q},\nu)$, $n_{\mathbf{q}\nu}$ is the Bose-Einstein phonon occupation number, and the sums run over all phonons with $\omega_{\mathbf{q}\nu} \leq \omega$.

All simulated spectra include the energy-dependent instrumental resolution.

\subsection*{Force constant parametrization}\label{sec:fc_parameter}

We adopted an orthogonal unit cell (Cmcm oxygen sublattice) with $a=4.51$~\AA{}, $b=7.82$~\AA{}, and $c=7.36$~\AA{}. The six force constants were optimized by least-squares fitting simultaneously to the experimental phonon band structure and thermal diffuse scattering:

\begin{table}[h]
    \centering
    \caption{Harmonic force constants optimized by least-squares fitting to experimental INS data.}
    \label{tab:force_constants}
    \begin{tabular}{lll}
        \hline
        Parameter & Value & Units \\
        \hline
        $k_\text{O-H}$ & 40.3 & eV \AA$^{-2}$ \\
        $k_{\text{O}\cdots \text{H}}$ & 1.02 & eV \AA$^{-2}$ \\
        $k_{\angle\text{O-H}\cdots \text{O}}$ & 5.20 & eV rad$^{-2}$ \\
        $k_{\angle\text{H-O}\cdots \text{H}}$ & 0.50 & eV rad$^{-2}$ \\
        $k_{\angle\text{H-O-H}}$ & 4.92 & eV rad$^{-2}$ \\
        $k_{\angle\text{H}\cdots \text{O}\cdots \text{H}}$ & 3.23 & eV rad$^{-2}$ \\
        \hline
    \end{tabular}
\end{table}

Detailed derivations of the force constant matrix block structures are provided in the Supplementary Information.

\subsection*{Mode-specific MSD calculations}\label{sec:msd}

Mode-specific mean-square displacements (MSDs) were computed from the dynamical matrix eigenvectors:  $\text{MSD}_b(\nu) = \langle|\delta b(\nu)|^2\rangle$ for bond stretching, where $\delta b(\nu)$ is the deviation from the equilibrium bond length; and $\text{MSD}_\alpha(\nu) = \langle r_{\alpha,1}r_{\alpha,2}|\delta \alpha(\nu)|^2\rangle$ for bond bending, where $\delta \alpha(\nu)$ is the angular deviation (in radians) and $r_{\alpha,1}, r_{\alpha,2}$ are the equilibrium lengths of the bonds forming angle $\alpha$. This prefactor converts the angular variance into spatially consistent units (\AA$^2$), making bending MSDs dimensionally comparable to the stretching MSDs.

\subsection*{Hydrogen configuration algorithms}\label{sec:algorithms}

\textbf{Random initialization with local projection.}
Initial hydrogen configurations were generated by locally projecting random Ising assignments onto the ice-rule manifold. For each oxygen site, the four associated Ising variables $\{\sigma_i\}$ were assigned at random, then projected onto the nearest valid two-in/two-out state by selecting the allowed configuration with the largest inner product (random choice in case of degeneracy). This procedure was iterated over all oxygen sites until all local ice-rule constraints $\sum_i \sigma_i = 0$ were satisfied, yielding a globally valid but statistically disordered configuration.

\textbf{Loop-flip equilibration.}
For large supercells, a smaller supercell ($10\times5\times5$) was first initialized using the local projection method and tiled to construct the target supercell ($20\times10\times10$). The tiling introduces artificial periodicity, which is removed by subsequent loop-flip equilibration; the resulting configurations are statistically indistinguishable from those obtained by direct initialization of the full supercell.

\subsection*{Thermodynamic simulation details}\label{sec:mcmc}

Markov-chain Monte Carlo (MCMC) simulations were performed on $20\times10\times10$ supercells using loop-flip dynamics, cooling linearly from $T = 3\Delta$ to $T = 0$ over 400,000 Monte Carlo steps. Trial moves consisted of collective hydrogen rearrangements along closed ``O-H$\cdots$O-H'' loops, accepted or rejected according to the Metropolis criterion based on $U(\mathbf{X})$. The 400,000-step protocol produces order-parameter trajectories and elastic diffuse scattering patterns indistinguishable from longer simulations, and $20\times10\times10$ supercells yield results consistent with larger systems ($24\times12\times12$).

The uncertainty in $\theta$ ($\pm15^\circ$) was determined from the range over which the Pearson correlation coefficient (Fig.~\ref{fig:4}d) drops visibly from its maximum. The uncertainty in $\Delta$ ($\pm15$~K) is the standard deviation of best-fit values across 20 independent MCMC cooling trajectories.

The simulated elastic diffuse scattering was computed as the static structure factor $S(\mathbf{Q}) = N\,e^{-\langle|\mathbf{Q}\cdot\mathbf{u}|^2\rangle}|\sum_i \bar{b}_i e^{i\mathbf{Q}\cdot\mathbf{r}_i}|^2$. Supercell convergence tests confirm that $20\times10\times10$ cells reproduce results from $24\times12\times12$ cells, with negligible finite-size effects on the elastic scattering.

The conformational entropy $S(X_\text{c}, X_\text{m})$ was estimated by counting configurations within the ice-rule manifold using an effective molecule number $N_\text{eff} = \frac{\ln(3/2)}{2\ln 3}N$, which rescales the naive combinatorial count to reproduce Pauling's residual entropy $S_{\omega\text{-Ih}} = k_\text{B}\ln(3/2)$ \cite{Pauling1935}. The full derivation is given in Supplementary Information.

\section*{End Notes}

\bmhead{Acknowledgements}

We wish to thank Rick Goyette, Saad Elorfi, Qiang Zhang, Nathan C Helton, Randall Sexton, Scott Tidwell, Robert Marrs, Chyan Duncan, Gary Lynn, Bobby Cross Jr., Andrew Parizzi, Silbino Vasquez, Douglas Engle, Colin Sarkis, Christian Balz, Alex Koldys (ORNL), Stephen Mills and Justin Link (Xavier University) for help with the sample handling and logistics. We wish to thank Takeshi Egami (ORNL) for insightful discussions. 

A portion of this research used resources at the Spallation Neutron Source, a DOE Office of Science User Facility operated by the Oak Ridge National Laboratory. The beam time was allocated to CORELLI and ARCS on proposal number IPTS-22517.1 and IPTS-33330.1.

\bmhead{Funding}
We acknowledge financial support by the Helmholtz Association of German Research Centers as well as Deutsche Forschungsgemeinschaft under Grants No. SFB 1143 and No. EXC 2147 ct.qmat. Travel funds for D.J.P.M. and J.L. were provided by the John Hauck Foundation and the Xavier University College of Arts and Sciences. The work of T.C., I.C.O. and D.A.T. was supported by the University of Tennessee Materials Research Science \& Engineering Center - The Center for Advanced Materials and Manufacturing - supported by the National Science Foundation under DMR No. 2309083.

\bmhead{Competing interests}
The authors have no competing interests.

\bmhead{Data availability}
Data and code supporting this study are available from the corresponding authors upon reasonable request during peer review and will be deposited in a public DOI-minting repository upon acceptance, with all data, simulation outputs, and materials underlying the study made publicly accessible at the time of publication.

\bmhead{Author contributions}
D.A.T., A.S., and D.J.P.M. designed and supervised the project. K.S., B.K., and D.J.P.M. prepared the samples. D.J.P.M., A.S., T.C., I.C.O., A.B., F.Y., J.L., and D.A.T. performed the neutron experiments. T.C., A.S., and I.C.O. analyzed the data. T.C. designed the models and performed the simulations. K.S., B.K., A.B., F.Y., D.L.A., and Z.J.M. provided resources. T.C., A.S., and Z.J.M. developed the software. T.C. wrote the manuscript and produced the figures with input from D.J.P.M., I.C.O., and D.A.T. All authors contributed to the discussion of the results and approved the final version of the manuscript.

\bmhead{Correspondence}
Correspondence and requests for materials should be addressed to T.C. and D.A.T.

\noindent

\clearpage

\clearpage
\clearpage

\section*{Supplementary Information}

\subsection*{Supplementary Note 1: Anomalous Phonon Modes from $^1$H$_2$O Frost}\label{secA1}

The phonon bands at 105~meV and 200~meV (Fig.~1b) observed in our measurements do not align with the simulated phonon dispersions for \mbox{$^2$H$_2$O}. This anomalous feature originates from \mbox{$^1$H} defects, which are easily detected in neutron scattering experiments due to the large incoherent scattering cross section of \mbox{$^1$H}. While both \mbox{$^1$H$_2$O} and \mbox{HDO} impurities introduce librational modes at 105~meV, their bending modes occur at 200~meV and 185~meV, respectively. Since no peak was observed at 185~meV in our data, we can rule out contributions from \mbox{HDO} and identify the impurity as \mbox{$^1$H$_2$O}. Furthermore, this identification suggests the $^1$H is not intrinsic to the crystal lattice; in a bulk \mbox{$^2$H$_2$O} environment, trace $^1$H would statistically favor the formation of \mbox{HDO} over \mbox{$^1$H$_2$O}. The dominance of pure \mbox{$^1$H$_2$O} modes implies that the $^1$H exists in a phase spatially separated from the \mbox{$^2$H$_2$O} lattice.

Evidence that these modes are extrinsic to the ice Ih structure comes from repeat measurements performed on the same sample after extended storage. We observed a relative increase in the 105~meV peak intensity compared to other phonon features over time. This enhancement supports the interpretation that the 105~meV mode arises from surface \mbox{$^1$H}-related impurities rather than intrinsic modes of the \mbox{$^2$H$_2$O} lattice. These findings strongly suggest that the impurity originates from polycrystalline \mbox{$^1$H$_2$O} frost forming on the sample surface during handling.

The peak intensities at 154~meV (\mbox{$^2$H$_2$O} bending modes) and 200~meV (\mbox{$^1$H$_2$O} bending modes) are nearly equal. Considering these modes are mostly incoherent, we use the incoherent scattering cross-sections of \mbox{$^2$H} (2.05~barn) and \mbox{$^1$H} (80.27~barn) to estimate that the \mbox{$^1$H$_2$O} frost contamination is approximately 2--3\%. This minor contamination does not affect the dispersive librational phonon features we observed because \mbox{$^1$H} contributes only weakly to the coherent scattering signal.

\subsection*{Supplementary Note 2: Stretching and Bending Interactions}\label{secA2}

We derive the force-constant matrices (FCMs) for stretching and bending interactions by first defining $\mathbf{r}_f$ and $\mathbf{r}_g$ as two arbitrary vectors that are linear combinations of the atomic coordinates $\mathbf{r}_i$ ($i=1\text{--}N$):
\[\begin{bmatrix}
         \mathbf{r}_f\\
         \mathbf{r}_g\\
     \end{bmatrix}=
\begin{bmatrix}
         f_1 & \cdots & f_N\\
         g_1 &\cdots & g_N\\         
     \end{bmatrix}
    \begin{bmatrix}
         \mathbf{r}_1\\
         \vdots \\
         \mathbf{r}_N\\
     \end{bmatrix}
 \]
so that $\mathbf{u}_f=\delta \mathbf{r}_f$ and $\mathbf{u}_g=\delta \mathbf{r}_g$ are linear combinations of atomic displacements $\mathbf{u}_i$ ($i=1\text{--}N$):
\[\begin{bmatrix}
         \mathbf{u}_f\\
         \mathbf{u}_g\\
     \end{bmatrix}=
\begin{bmatrix}
         f_1 & \cdots & f_N\\
         g_1 &\cdots & g_N\\         
     \end{bmatrix}
    \begin{bmatrix}
         \mathbf{u}_1\\
         \vdots \\
         \mathbf{u}_N\\
     \end{bmatrix}
 \]
For the stretching interactions, let $\mathbf{r}_f$ be the vector of bond $b$: $\mathbf{r}_f=\mathbf{b}$, $\mathbf{u}_f=\delta\mathbf{b}$.
$\delta\mathbf{b}$  can be decomposed into two components: $\delta\mathbf{b}^\parallel$, which is parallel to $\mathbf{b}$ and contributes to the stretching potential energy change, and $\delta\mathbf{b}^{\perp}$, which is perpendicular to $\mathbf{b}$ and only contributes to the reorientation of $\mathbf{b}$. The magnitude of the parallel component is given by $|\delta \mathbf{b}^\parallel|=\mathbf{u}_f\cdot \frac{\mathbf{r}_f}{r_f}$. The stretching potential energy is:
\[U_b=\frac{1}{2}k_b|\delta \mathbf{b}^\parallel|^2 =\frac{1}{2}k_b
\begin{bmatrix}
         \mathbf{u}_f^T\\
     \end{bmatrix}
  \begin{bmatrix}
         \frac{\mathbf{r}_f\mathbf{r}_f}{r_f^2}\\

     \end{bmatrix}
     \begin{bmatrix}
         \mathbf{u}_f \\
\end{bmatrix} \]
The corresponding FCM is:
  \[\mathbf{\Phi}_{\mathbf{b}=\mathbf{r}_f}=k_b\begin{bmatrix}
         f_1\\
         \vdots \\
         f_N\\         
     \end{bmatrix}
     \begin{bmatrix}
         \frac{\mathbf{r}_f\mathbf{r}_f}{r_f^2}\\ 
     \end{bmatrix}
\begin{bmatrix}
         f_1 & \cdots & f_N\\  
     \end{bmatrix}\]
For the bending interactions, let $\mathbf{r}_f$ and $\mathbf{r}_g$ be the two vectors of angle $\alpha$, and $\mathbf{n}$ be the unit vector perpendicular to both $\mathbf{r}_f$ and $\mathbf{r}_g$: $\mathbf{n}=\frac{\mathbf{r}_f \times \mathbf{r}_g}{|\mathbf{r}_f\times \mathbf{r}_g|}$. In the case where $\mathbf{r}_f\parallel \mathbf{r}_g$, $\mathbf{n}$ can be any unit vector perpendicular to $\mathbf{r}_f$, and the final FCM should be an average over all possible $\mathbf{n}$ (practically, we only need two vectors $\mathbf{n}_1$ and $\mathbf{n}_2$ that are perpendicular to each other). Define the angle vector $\boldsymbol{\alpha}=\alpha\mathbf{n}$. The change in the angle vector is given by: 
\[
\delta \boldsymbol{\alpha} = \mathbf{u}_f\times \frac{\mathbf{r}_f}{r_f^2}- \mathbf{u}_g\times \frac{\mathbf{r}_g}{r_g^2}
\]
$\delta\boldsymbol{\alpha}$ can be decomposed into two components: $\delta\boldsymbol{\alpha}^\parallel$, which is parallel to $\mathbf{n}$ and contributes to the change in bending potential energy, and $\delta\boldsymbol{\alpha}^{\perp}$, which is perpendicular to $\mathbf{n}$ and only contributes to the reorientation of $\boldsymbol{\alpha}$. The parallel component is:
\[
\delta \boldsymbol{\alpha}^\parallel = (\delta \boldsymbol{\alpha})\cdot \mathbf{n}=\mathbf{u}_f\cdot(\frac{\mathbf{r}_f}{r_f^2}\times \mathbf{n})-\mathbf{u}_g\cdot(\frac{\mathbf{r}_g}{r_g^2}\times \mathbf{n})
\]
If we define $\mathbf{w}_f=(\frac{\mathbf{r}_f}{r_f^2}\times \mathbf{n})$ and $\mathbf{w}_g=(\frac{\mathbf{r}_g}{r_g^2}\times \mathbf{n})$, then the bending potential energy is:
\[U_\alpha=\frac{1}{2}k_\alpha|\delta \boldsymbol{\alpha}^\parallel|^2 =\frac{1}{2}k_\alpha
\begin{bmatrix}
         \mathbf{u}_f^T & \mathbf{u}_g^T\\
     \end{bmatrix}
  \begin{bmatrix}
         \mathbf{w}_f\mathbf{w}_f^T & -\mathbf{w}_f\mathbf{w}_g^T\\
        -\mathbf{w}_g\mathbf{w}_f^T &  \mathbf{w}_g\mathbf{w}_g^T \\ 
     \end{bmatrix}
     \begin{bmatrix}
         \mathbf{u}_f \\
          \mathbf{u}_g \\
\end{bmatrix} \]
The corresponding FCM is:
  \[\mathbf{\Phi}_{\boldsymbol{\alpha}=\angle(\mathbf{r}_f,\mathbf{r}_g)}=k_\alpha\begin{bmatrix}
         f_1 & g_1\\
         \vdots & \vdots \\
         f_N & g_N\\         
     \end{bmatrix}
     \begin{bmatrix}
         \mathbf{w}_f\mathbf{w}_f^T & -\mathbf{w}_f\mathbf{w}_g^T\\
        -\mathbf{w}_g\mathbf{w}_f^T &  \mathbf{w}_g\mathbf{w}_g^T \\ 
     \end{bmatrix}
\begin{bmatrix}
         f_1 & \cdots & f_N\\
         g_1 & \cdots & g_N\\         
     \end{bmatrix}\]

\subsection*{Supplementary Note 3: Robustness and minimality of the six-parameter harmonic model}\label{secA3_robustness}

The central conclusions of the main text rely on a six-parameter harmonic description of the hydrogen-bond network. It is therefore important to show that this model is not an arbitrary choice, but rather the minimal experimentally identifiable representation that captures the measured lattice dynamics of ice Ih. 

Our retained model contains two 2-body stretching terms, for the short O-H covalent bond and the O$\cdots$H hydrogen bond, and four 3-body bending terms, for $\angle\mathrm{O-H\cdots O}$, $\angle\mathrm{H-O-H}$, $\angle\mathrm{H-O\cdots H}$ and $\angle\mathrm{H\cdots O\cdots H}$. Together these six force constants define a compact local basis for symmetry-allowed distortions of the hydrogen-bond network. The force-constant matrix for each stretching or bending interaction is constructed as described in Supplementary Note 2 and combined to form the full dynamical matrix.

To test whether the measured phonon spectra require additional harmonic terms, we explored a broader family of candidate models including extra 2-body stretching and 3-body bending interactions beyond the six retained terms. In particular, we examined medium-range pairwise stretching terms such as O$\cdots$O distortions as well as additional symmetry-allowed angular distortions. We find that most of these extra terms are not independently identifiable from the present neutron data: their effects on the calculated phonon dispersions and diffuse scattering are strongly correlated with those of the retained six force constants. For example, medium-range O$\cdots$O stretching largely renormalizes the effective stiffness already captured by the combined O-H and O$\cdots$H stretching channels, yielding little independent control over the measured phonon branches. Likewise, additional bending terms primarily project onto collective angular distortions already represented by the four retained angle-bending constants.

The six-parameter model is therefore best understood not as a unique microscopic decomposition of the intermolecular potential, but as the minimal independent harmonic basis that is both constrained by the data and sufficient to reproduce the observed anisotropic lattice dynamics. Larger models can introduce additional parameter freedom, but within the resolution of the present measurements they do not provide commensurate new spectroscopic information. Expanded parameterizations can yield marginal numerical improvements in selected parts of the spectrum, but these gains are small compared with the added parameter covariance and do not alter the qualitative physical conclusions.

This identifiability is reinforced by the fact that the experimental INS spectrum separates into well-resolved energy manifolds: acoustic and low-energy lattice modes below 39 meV, librational modes from 48 to 88 meV, intramolecular bending modes near 154 meV, and stretching modes near 308 meV. These distinct spectral sectors constrain different combinations of force constants. The high-energy bending and stretching manifolds primarily anchor the intramolecular stiffness scale, while the librational manifold constrains the hydrogen-bond network. Within the librational sector, the mode-resolved MSD analysis reported in the main text provides an additional, bond- and angle-resolved diagnostic of parameter sensitivity: the dominant coordinate for the soft-librational sector is $\angle\mathrm{H-O\cdots H}$, whereas the hard-librational sector is dominated by $\angle\mathrm{H\cdots O\cdots H}$. These experimentally resolved distinctions strongly restrict the relative bending-force hierarchy required of any acceptable model.

Most importantly, the physical conclusions of the manuscript are robust against reasonable extensions of the harmonic basis. Across all tested parameterizations that reproduce the measured phonon landscape with comparable quality, the same three qualitative results are preserved: (i) the anisotropic dispersive soft-lib modes arise naturally once the ice-rule-constrained hydrogen network is included; (ii) ice-rule constraints alone do not generate the dispersive hard-lib branch along $L$; and (iii) the broad librational bandwidth is controlled primarily by extended angular networks associated with $\angle\mathrm{H-O\cdots H}$ and $\angle\mathrm{H\cdots O\cdots H}$. The inference of additional hydrogen correlations beyond the ice rules therefore does not depend sensitively on the precise harmonic parametrization.

For these reasons, the six-parameter model used in the main text should be regarded as the most parsimonious experimentally constrained description of the observed phonon spectrum. It captures the full set of spectroscopic features relevant to the hidden-order interpretation while avoiding the parameter covariance and loss of identifiability introduced by larger harmonic models.

\subsection*{Supplementary Note 4: Estimation of Conformational Entropy}\label{secA4_entropy}

In ice Ih, each pair of nearest neighbor water molecules can adopt one of three possible geometric configurations (Fig.~3a).  If we ignore additional constraints, the total number of configurations is  
\[
W_\text{naive} = 3^{2N},
\]
where $N$ is the number of water molecules and $2N$ is the number of nearest neighbor molecule pairs.  Among them, $3N/2$ are center-symmetric and $N/2$ are mirror-symmetric. 

For a given order parameter $(X_{\text{c}}, X_{\text{m}})$, the number of configurations can be expressed as
\[
W(X_{\text{c}},X_{\text{m}};N)=
\binom {\frac{3}{2}N} {\frac{X_{\text{c}}}{2} N}
\binom {\frac{1}{2}N} {\frac{X_{\text{m}}}{2} N} 
2^{\left(2-\frac{X_{\text{c}}}{2}-\frac{X_{\text{m}}}{2}\right) N},
\]
and $\sum_{X_\text{c}, X_\text{m}} W(X_\text{c}, X_\text{m};N) = 3^{2N}$.  

However, the real number of configurations is reduced by constraints such as the ice rules. According to Pauling's estimate \cite{Pauling1935}, the total number of allowed hydrogen configurations is approximately  
\[
W_\text{Pauling} = (3/2)^N.
\]
To account for this reduction, we introduce an effective number of molecules
\[
N_\text{eff} = \frac{\ln(3/2)}{2\ln 3} \, N,
\]
so that $3^{2 N_\text{eff}} = (3/2)^N$.  

The corresponding conformational entropy per molecule is then
\[
S(X_{\text{c}},X_{\text{m}}) = \frac{k_\text{B}}{N} \ln W(X_{\text{c}},X_{\text{m}};N_\text{eff}).
\]
Using Stirling's approximation,
\[
\ln(x!) \approx \phi(x) - x,
\]
where $\phi(x)=x\ln(x)$, we can write the logarithm of a combination as
\[
\ln \binom{x}{y} \approx \phi(x) - \phi(y) - \phi(x-y).
\]
Then, for the number of configurations with the Pauling correction, we obtain
\[
S(X_\text{c}, X_\text{m}) \approx k_\text{B} 
\frac{\ln(3/2)}{4 \ln 3} \Biggl[
\phi(3)-\phi(X_\text{c})-\phi(X_\text{m})
- 2\phi\Bigl(\frac{3-X_\text{c}}{2}\Bigr) 
-2 \phi\Bigl(\frac{1-X_\text{m}}{2}\Bigr)
\Biggr].
\]

\section*{Supplementary Video Legends}

\textbf{Supplementary Video 1: Dynamics of hidden order formation.} 
Time evolution of the hydrogen sublattice during simulated cooling, corresponding to Fig.~4f. The animation visualizes the emergence and growth of polarized armchair chains (HA-chains) along the crystallographic $c$-axis. Chain color encodes the sign of the local polarization, while reduced opacity indicates mixed or weakly polarized configurations. The video illustrates the kinetic competition between domain growth and boundary frustration that leads to a mosaic of nanoscale polar domains.

\backmatter
\bibliography{bibliography}

\end{document}